\documentclass{jfm}
\usepackage{graphicx}
\usepackage{dblfloatfix} 
\usepackage{epstopdf, epsfig}
\usepackage{subfig}
\usepackage{caption}
\usepackage{graphicx}
\graphicspath{{figures/}}
\usepackage{float}
\usepackage{hyperref}
\usepackage{amssymb}
\usepackage{bm}
\hypersetup{
    colorlinks = true,
    linkcolor= blue,
    citecolor=blue,
    urlcolor=magenta
    }
    \usepackage{mathtools}
\usepackage{tabularx}
\shorttitle{Viscoelastic fluid flow over a transversely oscillating cylinder}

\title{Significant influence of fluid viscoelasticity on flow dynamics past an oscillating cylinder}

\author{Faheem Hamid\aff{1}
  \and C. Sasmal\aff{1}
  \corresp{\email{csasmal@iitrpr.ac.in}},
}

\affiliation{\aff{1}Department of Chemical Engineering, Indian Institute of Technology Ropar, Rupnagar, Punjab, India-140001.
}
\begin{document}
\maketitle
\begin{abstract}
This study presents a numerical investigation of how fluid viscoelasticity influences the flow dynamics past a transversely forced oscillating cylinder in the laminar vortex shedding regime at a fixed Reynolds number of $Re = 100$. In particular, we examine how fluid viscoelasticity influences the boundary between the lock-in and no lock-in zones and the associated wake topology compared to that seen in a simple Newtonian fluid. All in all, we find that the fluid viscoelasticity facilitates the synchronization of the vortex street with the cylinder motion at lower oscillation frequencies than that required for a Newtonian fluid. Consequently, the boundary of the lock-in region for a viscoelastic fluid differs from the Newtonian one and broadens in the non-dimensional cylinder oscillation amplitude and frequency plane. Furthermore, we propose that excess strain generated due to the stretching of polymer molecules in a viscoelastic fluid results in a marked difference in the wake structure from that seen in a Newtonian fluid. For a Newtonian fluid, only '2S' (two single vortices) and 'P+S' (a pair and a single vortex) vortex shedding modes are detected in the primary synchronization region. However, a ’2P’(two pairs of vortices) vortex mode is detected for a viscoelastic fluid in this region. We also employ the data-driven dynamic mode decomposition (DMD) reduced order modeling technique to extract and analyze underlying coherent flow structures and their associated frequencies to understand the differences in the flow dynamics of Newtonian and viscoelastic fluids past an oscillating cylinder.
\end{abstract}

\section{Introduction}
The flow past a fixed cylinder is a widely studied problem in fluid mechanics, both from the pragmatic and fundamental points of view. This benchmark problem provides significant information on several aspects of flow past a bluff body, such as boundary layer separation and vortex shedding, hydrodynamic drag and lift forces, transition among several flow states, etc. The flow physics becomes more complex and interesting when the cylinder oscillates either in the stream-wise or span-wise direction of the free stream~\citep{griffin1974vortex}. Numerous studies have been carried out in the past on the flow past a forced oscillating cylinder due to its immense engineering applications in structural, offshore, and thermal power engineering applications. In particular, the flow past a sinusoidally oscillating cylinder has received considerable interest because of the difference in the wake formation and vortex shedding phenomena from that of a stationary cylinder~\citep{anagnostopoulos2000n}. This dissimilarity manifests in terms of the synchronization or the lock-in condition and an increased vortex strength, which eventually impact the lift and drag forces acting on the cylinder~\citep{kumar2013flow, bishop1964lift}. In the lock-in condition, the vortex shedding frequency diverges from the natural vortex shedding frequency corresponding to a stationary cylinder (known as the Strouhal frequency) and synchronizes with the imposed oscillating frequency of the cylinder~\citep{bishop1964li}. Numerous experimental~\citep{bishop1964li, koopmann1967vortex} and numerical~\citep{blackburn1999study, karniadakis1989frequency, anagnostopoulos2000numerical} investigations have been performed to understand and characterize the lock-in phenomenon for a wide range of Reynolds numbers~\citep{jones1968unsteady}, cylinder oscillation amplitude and frequency~\citep{koopmann1967vortex}. All these stated works aimed to determine the boundary between the lock-in and no lock-in conditions on the cylinder oscillation amplitude and frequency space. Also, sufficient information is now available on various flow aspects such as different wake structures (e.g., '2S'-two single vortices, '2P'-two pairs of vortices, and 'P+S'-a pair and a single vortices shed in one oscillation cycle)~\citep{williamson1988vortex, leontini2006wake}, mechanisms of vorticity production~\citep{blackburn1999study}, and hydrodynamic forces acting on the cylinder. 

\par However, to date, almost all of these studies have dealt with simple Newtonian fluids like water. On the other hand, most of the fluids encountered from our daily lives to several industrial settings exhibit various non-Newtonian behaviors, such as shear-thinning, shear-thickening, viscoplasticity, viscoelasticity, etc~\citep{chhabra2011non}. Despite this, almost no study is available on how the complex rheological behavior of a fluid could influence the flow dynamics past a forced oscillating cylinder, particularly the lock-in and no lock-in boundary and the associated vortex dynamics and hydrodynamic forces. Only very recently, Alam et al.~\citep{alam2021numerical} investigated the effect of the shear-thinning rheological behaviour and found a significant difference in the vortex structures depending upon the values of the cylinder oscillation amplitude and frequency. Viscoelasticity is also another non-Newtonian behaviour that is often seen in many complex fluids~\citep{bird1987dynamics}. Adding a minute amount of solid polymers, even in parts per million (ppm) quantity, into a Newtonian solvent like water dramatically changes the flow behaviour of the resulting polymer solution compared to that seen in water alone. This is due to the presence of non-linear elastic stresses in these fluids, originating from the stretching and relaxation of polymer molecules in a deformed flow field. To the best of our knowledge, how this fluid viscoelasticity would tend to influence the flow dynamics of a periodically forced oscillating cylinder is not investigated yet. The present study aims to fill that knowledge gap in the literature. In particular, this study investigates how one can alter the boundary between the lock-in and no lock-in regimes and the associated vortex dynamics in flows past a forced oscillating cylinder by inducing elasticity in a viscous fluid. This could be useful in many practical applications dealing with oscillating structures and viscous fluids. We also employ one of the widely used reduced order modeling technique, the dynamic mode decomposition (DMD) technique, in the present study for visualizing and better understanding the difference between the viscoelastic and Newtonian fluid flow dynamics, especially in the lock-in and no lock-in regimes.
\begin{figure}
    \centering
    \includegraphics[trim=0cm 0cm 0cm 0cm,clip,width=12cm]{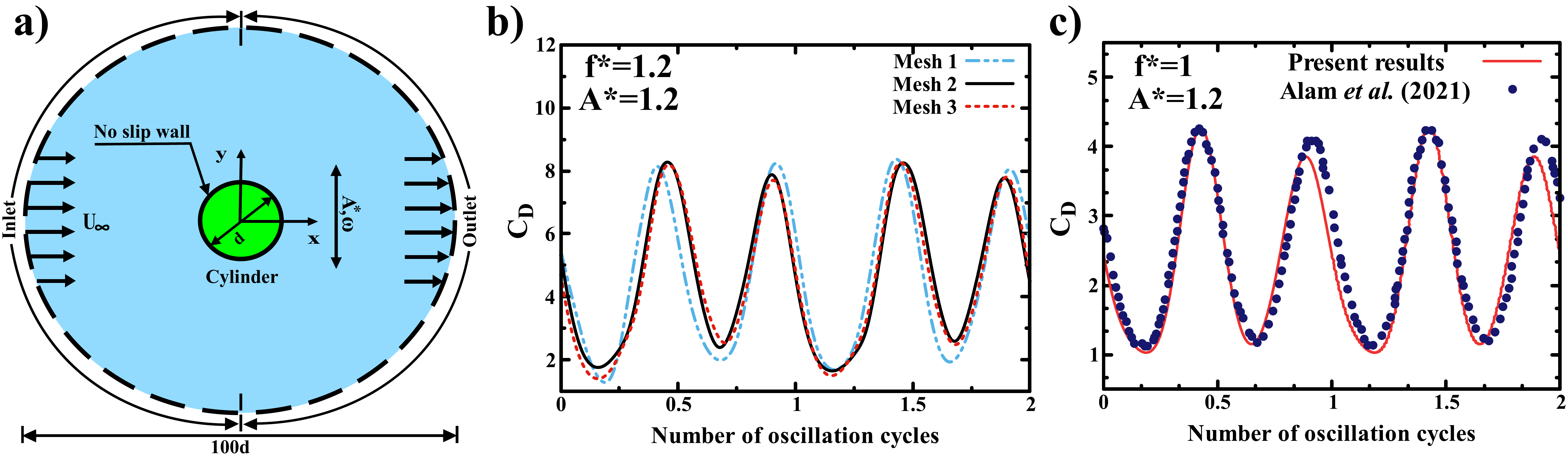}
    \caption{(a) Schematic of the present problem, (b) temporal variation of drag coefficients ($C_{D}$) for three different mesh densities, and (c) comparison of drag coefficient between the present results with that of Alam et al.~\citep{alam2021numerical}.}
    \label{fig:Schematic of the problem 1}
\end{figure}
\section{Problem statement and governing equations}\label{sec:PSGE}
We investigate a two-dimensional, laminar, and unbounded flow of viscoelastic fluid past a periodically forced oscillating cylinder of diameter $d$, as shown schematically in figure~\ref{fig:Schematic of the problem 1}(a). The cylinder is placed at the center of a fictitious circular domain of viscoelastic fluid of diameter $100d$, wherein the fluid enters the domain with a uniform velocity of $U_{\infty}$. The cylinder is considered to be infinitely long in the z-direction so that no gradient exists in this direction, i.e.,$ \dfrac{\partial{()}}{\partial{z^{*}}}=0$ and also there is no flow in this direction, i.e., $u_{z}^{*}=0$. Furthermore, the cylinder is forcefully oscillated by applying a harmonic oscillation with an amplitude of $A^*$ (normalized) and a frequency of $f_{y}^{*}$ such that the normalized displacement $(Y)$ of the cylinder with respect to time $(t^{*})$ is given by $Y=A^*$Sin$(2\pi f_{y}^{*}t^{*})$. Here, the cylinder oscillation amplitude and frequency are normalized with the cylinder diameter ($A^{*}=\frac{A}{d}$) and the Strouhal frequency ($f^{*}=\frac{f_{y}^{*}}{f_{st}}$), respectively.  
{\noindent The incompressible flow of the viscoelastic fluid, whose rheological behaviour is mimicked with the Oldroyd-B constitutive equation~\citep{oldroyd1950formulation}, is governed by the following dimensionless equations:}\newline
Continuity equation:
\begin{equation}
             \nabla\cdot\textbf{u}=0  
\end{equation}
\begin{flushleft}Cauchy momentum equation:\end{flushleft}
\begin{equation}
\dfrac{\partial{\textbf{u}}}{\partial{t}}+ \textbf{u}\cdot\nabla{\textbf{u}} = -\nabla{p} + \frac{\beta}{Re}\nabla\cdot\boldsymbol{\tau_{s}}+\frac{(1-\beta)}{Wi}\frac{1}{Re}\nabla \cdot\boldsymbol{\tau_{p}}
\end{equation}
\begin{flushleft}Oldroyd-B constitutive equation:\end{flushleft}
\begin{equation}
  Wi \left( \frac{\partial \bm{\tau_{p}}}{\partial t} + \bm{u} \cdot \nabla \bm{\tau_{p}} - \bm{\tau_{p}} \cdot \nabla \bm{u} - \left( \nabla \bm{u} \right)^{T} \cdot \bm{\tau_{p}} \right) + \bm{\tau_{p}} = \left(1 - \beta\right) \left( \nabla \bm{u} + \nabla \bm{u}^{T}\right)
\end{equation}
The above equations have been non-dimensionalized using parameters $d$,\,\,$\frac{d} U_{\infty}$,\,\,$U_{\infty}$,\,\,$\rho\,U_{\infty}^2$,\,\, \\$\frac{\eta_{0}\,U_{\infty}}{d}$ for length ($\bm{x}^{*}$), time ($t^{*}$), velocity ($\textbf{u}^*$), pressure ($p^*$), and stress tensor ($\boldsymbol{\tau}^*$), respectively. Here, $\rho$ is the fluid density, $\eta_{0}$ is the zero-shear rate viscosity of the polymer solution given by the sum of the solvent viscosity, and the contribution of the polymer, i.e., $\eta_{0} = \eta_{s} +\eta_{p}$. The total deviatoric stress tensor in the fluid is a linear combination of the viscous term $\boldsymbol{\tau_{s}^*}$ (the solvent contribution) and the viscoelastic extra-stress term $\boldsymbol{\tau_{p}^*}$ (the polymer contribution). The Newtonian solvent contribution to the stress tensor is given by $\boldsymbol{\tau_{s}}^* = 2\eta_{s}\textbf{D}^*$, where $\textbf{D}^*$ is the rate of deformation tensor given by  $\frac{1}{2} \left[ \left(\nabla^*{\textbf{u}^*} \right) + \left( \nabla^*{\textbf{u}^*} \right)^{T}\right]$  and $\eta_{s}$ is the viscosity of the Newtonian solvent. It can be seen that the present flow is governed by three dimensionless numbers, namely, Reynolds number defined by \,$ Re=\frac{\rho\,U_{\infty}\,d}{\eta_{0}}$, Weissenberg number defined by  $Wi=\frac{\lambda\,U_{\infty}}{d}$, and polymer viscosity ratio defined as $ \beta= \frac{\eta_{s}}{\eta_{s}+\eta_{p}}$. Here, $\lambda$ is the characteristic relaxation time of polymer molecules. Apart from these, normalized cylinder oscillation amplitude $(A^{*})$ and frequency $(f^*)$ are two additional dimensionless numbers that will govern the present flow. The present investigation is carried out at fixed values of $Re = 100$, $Wi = 2$, $\beta = 0.5$ and for a range of values of  $0.4\leq\, A^{*}\,\leq1.2$ and $0.6\leq\,f^{*}\,\leq1.2$. Additionally, simulations for a Newtonian fluid have been carried out to directly compare the flow dynamics in viscoelastic fluids under the same conditions.  


\section{Numerical method and validation}
A detailed description of the numerical method employed in the present study is available elsewhere~\citep{hamid2022dynamic,khan2020flow}, and hence, it is not repeated here for the sake of brevity. Broadly, we have utilized the finite volume method (FVM) based open source code OpenFoam coupled with RheoTool~\citep{rheoTool} to solve the governing mass, momentum, and Oldroyd-B viscoelastic constitutive equations. Details of the discretization techniques applied to various terms of the governing equations are provided in our earlier study~\citep{hamid2022dynamic}. Additionally, an automatic mesh motion scheme (dynamicFvMesh) has been used to treat the fluid-structure interaction, wherein the mesh at the outer boundary is fixed. At the same time, it deforms in the vicinity of the cylinder due to its motion. A grid independence study was conducted by making three different grids consisting of regular hexahedral elements, namely, Mesh 1(54000), Mesh 2 (102000), and Mesh 3 (206800), to choose an optimal mesh density. Figure~\ref{fig:Schematic of the problem 1}(b) shows the temporal variation of the drag coefficient $C_{D}$ at extreme values of oscillation parameters used in the present study. After carefully inspecting the results obtained with three different mesh densities, Mesh 2 was chosen for the present study as  Mesh 2 (102000) and Mesh 3 (206800) values show a nearly perfect match. Furthermore, a time step size of $\Delta t=0.0001\frac{d}{U_{\infty}}$ was chosen to carry out all the simulations after performing a rigorous time independence study. Some validation studies have also been carried out to establish the accuracy and reliability of the present numerical setup. Figure~\ref{fig:Schematic of the problem 1}(c) shows the comparison of the drag coefficient with the results of Alam et al.~\citep{alam2021numerical}. A very good correspondence can be seen between the two results. Finally, the following boundary conditions were employed to facilitate the present simulations. At the inlet, $u_{x}=1 $ and $u_{y}=0$ are used. Furthermore, the pressure gradient and the extra stresses due to polymeric contribution are set to zero at this boundary. At the Outlet, the pressure is set to zero value, and the Neumann boundary condition is used for the rest of the variables. At the cylinder surface, the standard no-slip boundary condition is applied along with a zero gradient for the pressure. The polymeric stresses are extrapolated linearly onto this surface. 
\par The DMD analysis has been carried out using the algorithm proposed by Schmid~\citep{schmid2011}, which is also detailed in our earlier study~\citep{hamid2022dynamic}. Briefly put, temporally equispaced (with 0.1s interval) snapshots ( a total of $N = 501$) of the vorticity field in the unsteady periodic regime ($t\geq 500$) are vectorized and assembled to form a matrix $S_{1}^{N}=\{s_{j}\}_{j=1}^N$. DMD assumes a linear map $\mathbf{M}$ that connects the consecutive vorticity fields as $s_{j+1}=\mathbf{M}s_{j}$. Therefore, the system can be arranged as $S_{2}^{N}=\mathbf{M}S_{1}^{N-1}= \mathbf{C}S_{1}^{N-1}+ \mathbf{r}$. Here, $\mathbf{r}$ is the residual, which is minimized to compute the eigenvalues (Ritz Values denoted by $\lambda_{j}$) and eigenvectors (DMD modes denoted by $\phi_{j}$) of $\mathbf{C}$ by singular value decomposition (SVD). The DMD modes capture the spatial coherent features of the flow field, whereas the Ritz values ($\lambda_{j}$) provide information about the growth rate ($\sigma_{j} = Re (log(\lambda_{j}/t))$) and frequency ($St_{j} = Im (log(\lambda_{j})/2\pi\,t)$) of these modes. The energy contribution of each mode is quantified using the mode amplitude $b_{j}$, which is obtained by $\mathbf{b}=\mathbf{\Phi}^{\dag}s_{1}$. Here, $\mathbf{b}$ is the matrix of amplitude vectors,  $\Phi^{\dag}$ denotes the Moore-Penrose pseudoinverse of mode matrix, and $s_{1}$ is the initial snapshot vector.

\section{Results and discussion}\label{sec:Results_and_discussion }
\begin{figure}
    \centering
    \includegraphics[trim=0cm 0cm 0cm 0cm,clip,width=12cm]{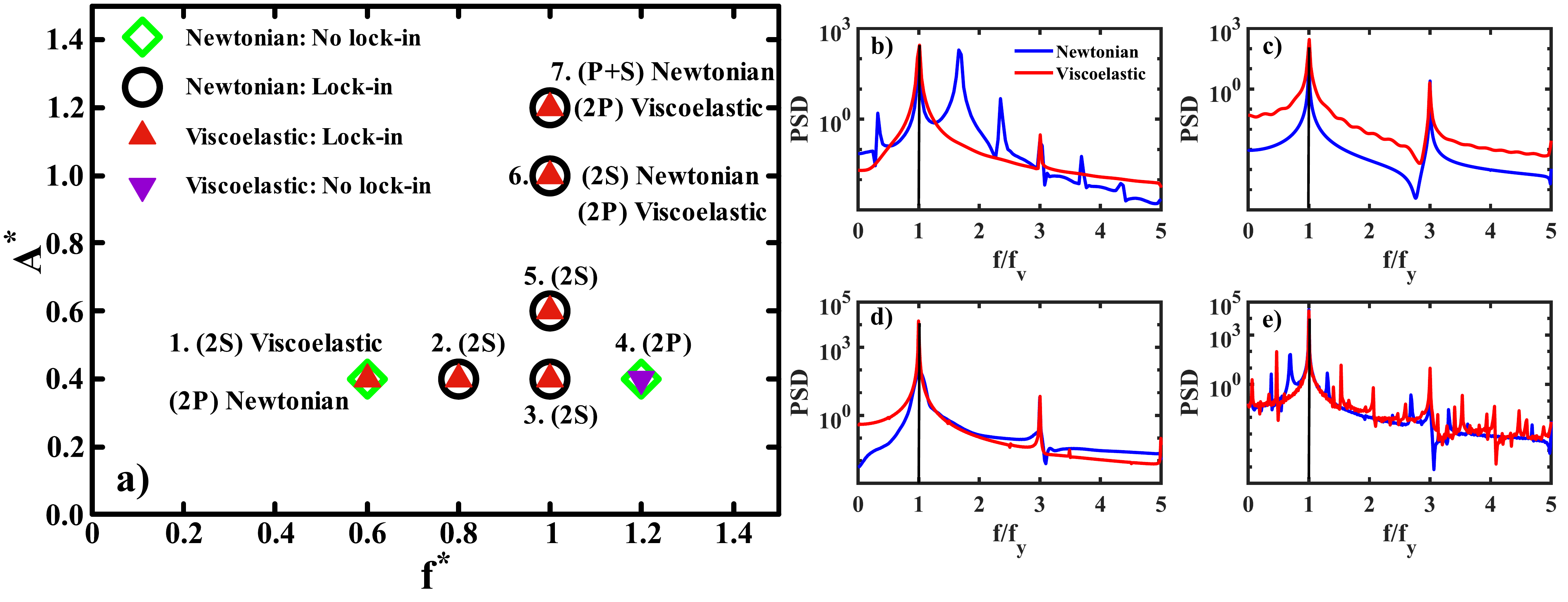}
    \caption{ (a) Map illustrating the lock-in and no lock-in states for viscoelastic  and Newtonian fluids at selected points in the non-dimensional amplitude ($A^{*}$) and frequency ($f^*$) plane. Power spectrum density plots of the lift coefficients at $A^{*}=0.4$ and frequency of: b) $f^{*}=0.6$, c) $f^{*}=0.8$, d) $f^{*}=1$, e) $f^{*}=1.2$.}
    \label{fig:2}
\end{figure}

\begin{figure}
    \centering
    \includegraphics[trim=0cm 0cm 0cm 0cm,clip,width=14cm]{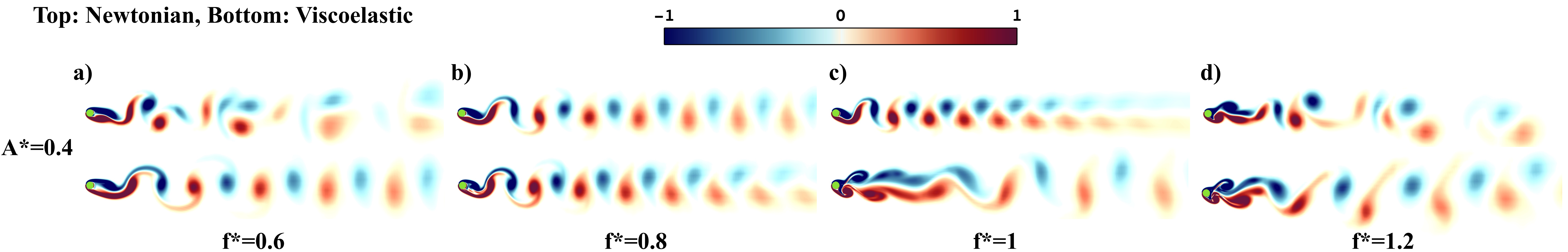}
    \caption{Vortex structures in Newtonian and viscoelastic fluids corresponding to points 1, 2, 3, and 4 in the map as shown in figure~\ref{fig:2}(a).}
    \label{fig:3}
\end{figure}

We start the discussion by presenting the comparison of the lock-in and no lock-in states seen between viscoelastic and Newtonian fluids at selected combinations of $A^{*}$ and $f^{*}$. Subsequently, for the same combinations, we illustrate in detail how the associated vortex dynamics differs in viscoelastic and Newtonian fluids. It is followed by the dynamic mode decomposition (DMD) analysis of two cases when i) both Newtonian and Viscoelastic fluids are in the lock-in condition ($A^{*}=0.4$ and $f^{*} = 0.8$), and ii) viscoelastic fluid deviates from the Newtonian one (no lock-in) and exhibits the lock-in condition ($A^{*}=0.4$ and $f^{*} = 0.6$). In the present study, we have chosen seven sets of $A^{*}$ and $f^{*}$ based on a recent study by Alam et al.~\citep{alam2021numerical} for elucidating the effect of the fluid viscoelasticity on the lock-in and no lock-in behaviours. First, the non-dimensional frequency-amplitude map exhibiting the lock-in and no lock-in behaviors of Newtonian and viscoelastic fluids is presented in figure~\ref{fig:2}(a). From this plot, one can clearly observe that at a combination of $A^{*}=0.4$ and $f^{*} = 0.6$, viscoelastic fluid exhibits the lock-in state, whereas Newtonian fluid is in the no lock-in state. However, as the value of $f^{*}$ increases to 0.8 or 1 at the same value of $A^{*}=0.4$, both fluids display the lock-in behavior. On further increasing $f^{*}$ to 1.2, both fluids move to the no lock-in zone. This suggests a stark difference in the synchronization behaviour of the wake structures for the value of $f^{*} < 0.8$ between viscoelastic and Newtonian fluids. This information is obtained from the power spectral density (PSD) plot of the temporal variation of the lift coefficient presented in figure~\ref{fig:2}(b-e). As can be seen, for $f^{*}< 1$, the viscoelastic fluid in both cases (b corresponding to point 1 in the map, and c corresponding to point 2 in the map) has a dominant peak at the cylinder oscillation frequency and another peak at its integer multiple, thereby fulfilling the criteria of the lock-in condition. Contrary to this, at a low frequency of $f^{*}=0.6$, the power spectrum of the lift coefficient of Newtonian fluids has several peaks at non-integer multiples of the cylinder oscillation frequency, indicating a no lock-in behavior of the wake. Furthermore, the lock-in condition of the wake for an Oldroyd-B fluid at $f^{*}=0.6$ is easily discernible from the vortex patterns shown in figure~\ref{fig:3}(a). Here, the wake in a Newtonian fluid is in the non-synchronized state, with a pair of vortices (2P) being shed in one oscillation cycle. On the other hand, for an Oldroyd-B fluid, two single vortices (2S) are shed into the wake, which is again synchronized with the cylinder oscillation. Moreover, these vortices are found to be stretched in the stream-wise direction with a connecting region of vorticity between the two subsequent vortices in the near wake zone. On further moving to higher oscillation frequencies, i.e., $f^{*}=0.8$ and $f^{*}=1$, both fluids exhibit the lock-in behavior (see figure~\ref{fig:2}(c) and (d) for the corresponding PSD plots) with 2S mode of vortex shedding. However, the vortex patterns are significantly different from each other. While for a Newtonian fluid, the wake at $f^{*}=0.8$ resembles the von-Karman vortex street, the vortices in the far wake organize into rows for an Oldroyd-B fluid. Moreover, this trend of a viscoelastic fluid is replicated by the Newtonian fluid at $f^{*}=1$, although with a higher rate of vortex dissipation. At this point ($f^{*}=1$), vortices in an Oldroyd-B fluid display substantial stretching in the downstream direction before being shed with higher lateral spacing. Much similar trend is observed at the frequency ratio of $f^{*}=1.2$  where the PSD plots (figure~\ref{fig:2}(e)) show a number of small amplitude peaks, which are again more pronounced in an Oldroyd-B fluid than in the Newtonian one. In both fluids, the non-synchronized vortices advect behind the cylinder with a 2P configuration; however, the Oldroyd-B vortices are noticeably stronger than the Newtonian ones. 
\par
\begin{figure}
    \centering
    \includegraphics[trim=0cm 0cm 0cm 0cm,clip,width=14cm]{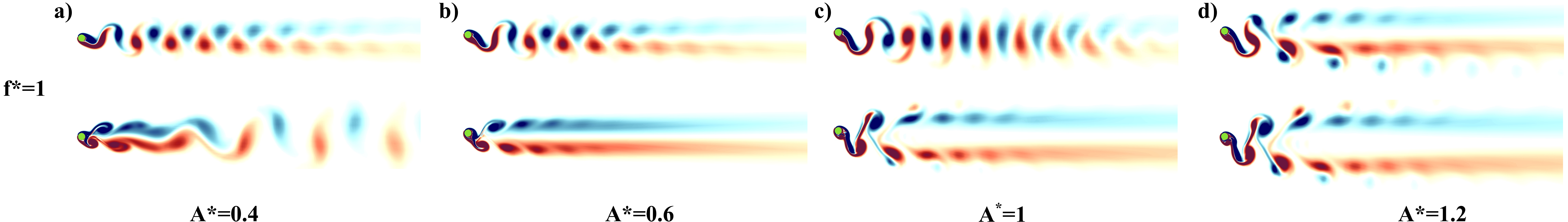}
    \caption{Vortex structures in Newtonian and viscoelastic fluids corresponding to points 3, 5, 6, and 7 in the map as shown in figure~\ref{fig:2}(a).}
    \label{fig:4}
\end{figure}

 On the other hand, at a fixed frequency of $f^{*}=1$, as the oscillation amplitude increases from $A^{*}=0.4$ to $A^{*}=1.2$, both Newtonian and viscoelastic fluids stay in the lock-in state. However, a notable difference is again seen in the wake dynamics, as illustrated in figure~\ref{fig:4}. First, at $A^{*}=0.6$, for an Oldroyd-B fluid, the vortices are shed with '2S' mode but in parallel rows from the near wake itself, whereas this partition occurs at some distance for a Newtonian fluid. With the further increment in the amplitude ratio to $A^{*}=1$, a notable difference appears in the wake topology. While for a Newtonian fluid, the wake still remains in '2S' mode, the wake shifts to '2P' mode for a viscoelastic fluid. This is a significant deviation from the Newtonian behavior where only '2S' or 'P+S' modes have been observed in the primary synchronization region at multiple Reynolds numbers in earlier studies~\citep{leontini2006wake}. A possible explanation for this, already provided by Govardhan and Willianmson~\citep{govardhan2000modes}, is that the single vortices in the near wake split due to excess strain resulting in the pairing of vortices. Since polymer molecules also undergo considerable stretching in the cylinder vicinity in a viscoelastic fluid, the excess straining effect enhances the pairing of the vortices at lower cylinder oscillation amplitude and frequency values than the Newtonian fluid. For a Newtonian fluid, a synchronized 'P+S' wake state is observed at a higher amplitude ratio of $A^{*}=1.2$, as reported by many earlier studies.  
\par
\begin{figure}
    \centering
    \includegraphics[trim=0cm 0cm 0cm 0cm,clip,width=12cm]{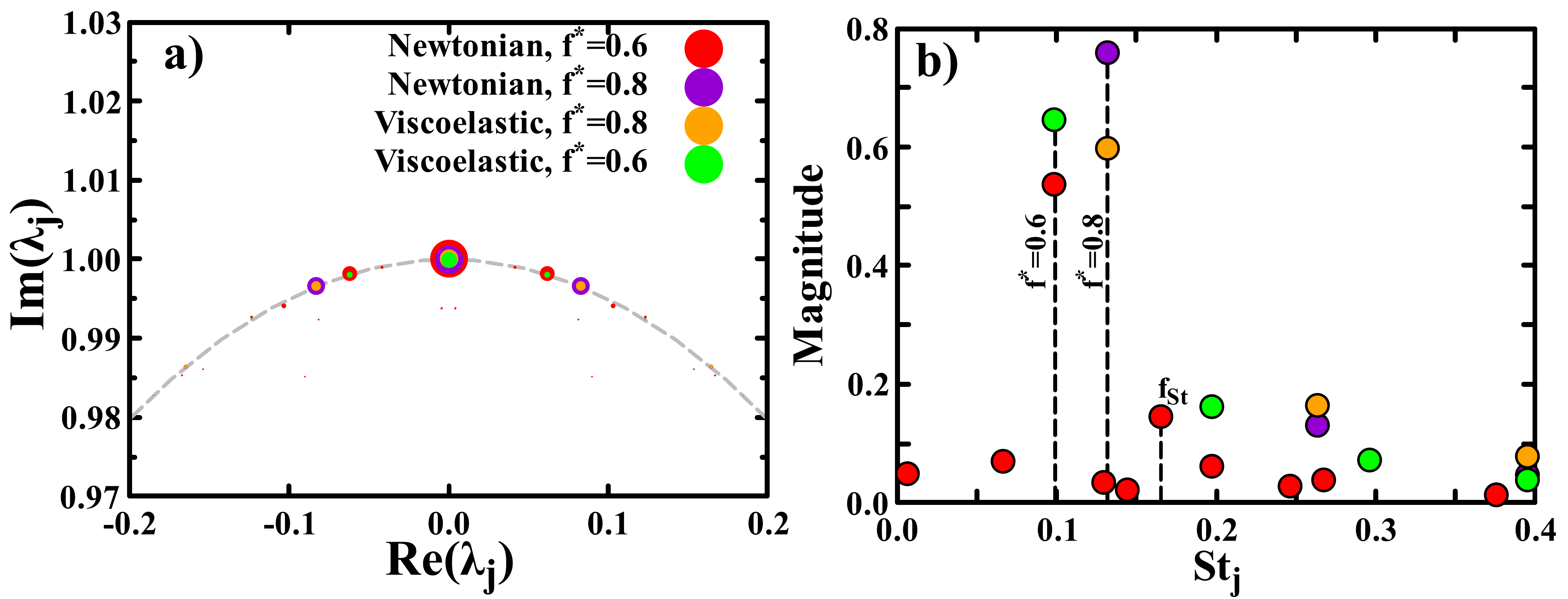}
    \caption{(a) The Ritz values ($\lambda_{j}$) both for Newtonian and viscoelastic fluids at $A^{*}=0.4$ and different values of $f^{*}$, (b) the normalized magnitudes corresponding to the most dominant modes (excluding the mean mode) are plotted against the non-dimensional frequency of each mode. Fundamental frequencies in the fluid flow are denoted by the dashed lines. The rest of the frequencies are higher harmonics of these fundamental frequencies in the flow.}
    \label{fig:5}
\end{figure}

 Next, we utilize the dynamic mode decomposition (DMD) technique to analyze the differences in the coherent flow structures of both fluids at $f^{*}=0.6$, where they exhibit different synchronization behavior. We also perform the DMD analysis at $f^{*}=0.8$ as well, where both fluids are in the lock-in state. DMD extracts the underlying structures of dynamic relevance from the global flow field data and their associated frequencies prevalent in the domain. This will aid in visualizing the competition between the cylinder oscillation and the natural vortex shedding frequencies in the wake. First, the Ritz values of all the cases under consideration are plotted in figure~\ref{fig:5}(a). Irrespective of the fluid type and oscillation parameters, most of these neutrally stable values are clustered around the unit circle $|\lambda_j|=1$ with a few strongly damped values lying inside it. It denotes the convergence towards a linear representation of the nonlinear flow. Also, these values are symmetrical with respect to the real axis due to real flow field data. In our analysis, the sorting of dominant modes is done based on the amplitude ($b_j$), which is determined by projecting the modes back to the original data sequence. The modes with larger projections are more significant and are depicted by larger sizes in figure~\ref{fig:5}(a). 
The mode with zero imaginary Ritz value is shown by the largest circle for all fluids and captures the mean vorticity field. For other modes, the normalized magnitudes are plotted against the associated frequencies ($St_{j}$) in figure~\ref{fig:5}(b). From this plot, it is obvious that only a few modes capture the maximum flow energy in all four scenarios, which is expected as the flow is periodic. To begin, for the lock-in case (i.e., $f^{*}=0.8$), both Oldroyd-B and Newtonian fluids have a pronounced peak at the cylinder oscillation frequency. The rest of the frequencies are higher harmonics of this frequency and contribute negligibly to the data sequence as reflected from their amplitudes. Moreover, the natural vortex shedding frequency ($f_{st}$) of both fluids does not appear in this spectrum. It further confirms that the vortex shedding synchronizes with the cylinder oscillation in the lock-in condition. Therefore, the cylinder oscillation dominates while the natural shedding frequency perishes in the flow field. 

\begin{figure}
    \centering
    \includegraphics[trim=0cm 0cm 0cm 0cm,clip,width=12cm]{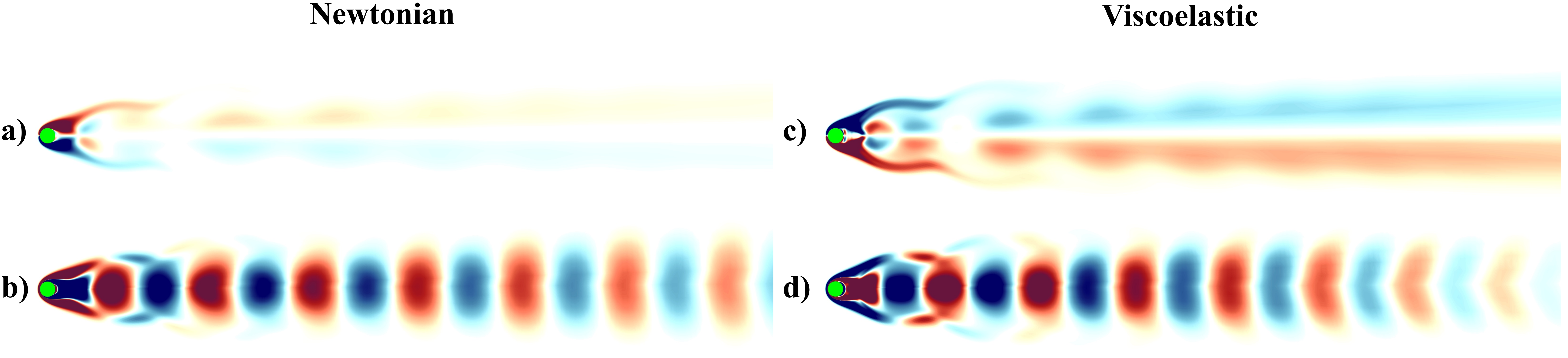}
    \caption{Visualization of the DMD modes of the vorticity ﬁelds both for Newtonian and viscoelastic fluids at the lock-in state in the non-dimensional amplitude-frequency plane ($A^{*}=0.4$ and $f^{*}=0.8$). (a, c) mean mode, (b, d) mode 1.}
    \label{fig:6}
\end{figure}

\begin{figure}
    \centering
    \includegraphics[trim=0cm 0cm 0cm 0cm,clip,width=12cm]{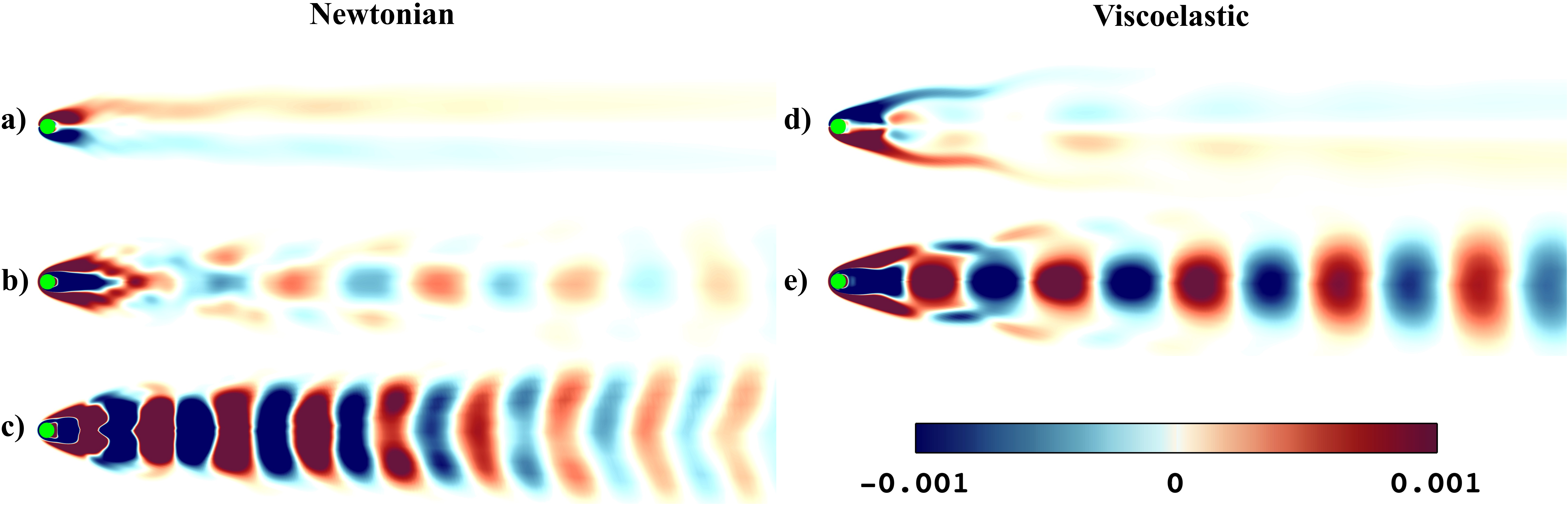}
    \caption{Visualization of the DMD modes of the vorticity ﬁeld both for Newtonian (no lock-in) and viscoelastic (lock-in) fluids at the deviation point in the non-dimensional amplitude-frequency plane ($A^{*}=0.4$ and $f^{*}=0.6$). (a, d) mean mode, (b, e) mode 1, (c) mode 2.}
    \label{fig:7}
\end{figure}

Figure~\ref{fig:6} shows the real parts of the DMD modes corresponding to the mean flow and the cylinder oscillation frequency both for Newtonian and viscoelastic fluids. 
In both fluids, the mean mode ($St_{j}=0$) comprises two shear layer-type structures arising behind the cylinder and extending symmetrically into the wake. However, these structures are more prominent for a viscoelastic fluid and extend further into the wake. Mode 1 provides typical structures to capture the vortex shedding phenomenon and its associated frequency. For both fluids, the bubble-like structures convect in an anti-symmetric fashion away from the cylinder with its oscillation frequency. However, there is a marked difference in the far wake structures between the two fluids. Contrary to the flow of viscoelastic fluids past a stationary cylinder~\citep{hamid2022dynamic} where the strength of these structures is higher than Newtonian fluids, the opposite trend is observed here. This is because, in a Newtonian fluid, single vortices ('S') sustain to a greater distance in the wake than in a viscoelastic fluid. In this fluid, the vortex street is partitioned into two rows in the far wake (see figure~\ref{fig:3}), which is successfully captured by the DMD structures. Figure~\ref{fig:7} depicts the modes for Newtonian no lock-in and viscoelastic lock-in cases at $A^{*}=0.4$ and $f^{*}=0.6$. As seen before, in the lock-in condition, the DMD extracts only one dominant frequency in viscoelastic flows, i.e., the cylinder oscillation frequency (see figure~\ref{fig:5}(b)), and the other low magnitude frequencies are its higher harmonics. However, for the Newtonian no lock-in case, two fundamental frequencies are visible due to the non-synchronization in the vortex formation. Hence, from the flow field data, it is obvious that these two frequencies, i.e., the cylinder oscillation frequency and the vortex shedding frequency, compete in the flow domain. The associated flow structures at these frequencies are visualized in figure~\ref{fig:7}, where the mean mode for both fluids has similar structures as explained in the earlier case. Moving to mode 1 in a viscoelastic fluid, again, the same vortex-shedding structures appear; however, these resemble the DMD structures of the Newtonian lock-in case (see figure~\ref{fig:5}(b)), albeit more concentrated due to the viscoelastic effect. This observation of similarity in the DMD structures is also reinforced by the vortex topology at two points in figure~\ref{fig:3} (see bottom: $A^{*}=0.4$ and $f^{*}=0.6$, and top: $A^{*}=0.4$ and $f^{*}=0.8$), and also suggests that inducing viscoelasticity in a Newtonian fluid expedites the transition to the lock-in region for a transversely oscillating cylinder. Now looking at the DMD mode 1 for a Newtonian fluid, which corresponds to the cylinder oscillation frequency, the concentrated structures in the cylinder vicinity signify the dominance of forced oscillations in the near wake zone. These structures fade as we move downstream of the cylinder. Competing with this frequency, the anti-symmetric structures in mode 2 move with the natural shedding frequency and are more stretched in the span-wise direction as compared to the lock-in vortex shedding structures, see figure~\ref{fig:7}(c).  
\par
\begin{figure}
    \centering
    \includegraphics[trim=0cm 0cm 0cm 0cm,clip,width=12cm]{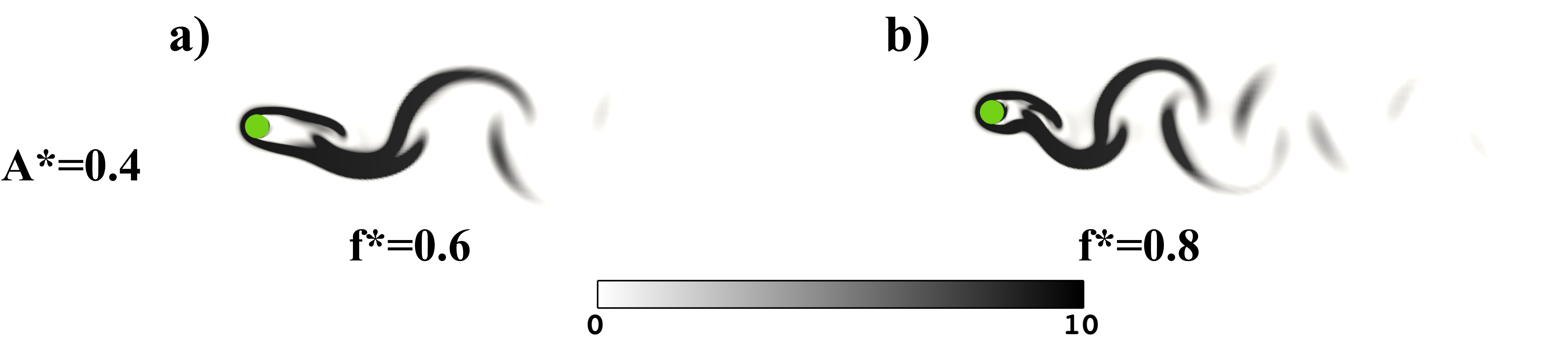}
    \caption{Stretching of polymer molecules in viscoelastic ﬂuid at $A^{*}=0.4$ and frequency ratios of $f^{*}=0.6$ and $f^{*}=0.8$.}
    \label{fig:8}
\end{figure}

 Finally, we propose that this deviation of viscoelastic fluid behavior from the Newtonian one is attributed to the stretching of the polymer molecules. We quantify and visualize this polymer stretching using the non-dimensional conformation tensor, figure~\ref{fig:8}. It can be seen that highly stretched strands of polymer molecules are formed in the cylinder vicinity, which further extend into the downstream wake. These strands suppress the vortex shedding as the polymer molecules sustain the stresses for a longer duration, which results in the elongation of the vortices (see figure~\ref{fig:3} $A^{*}=0.4$ and $f^{*}=0.6$). As a result, the inertial effects generated by the cylinder are not dissipated but sustained in the fluid. Therefore, the cylinder oscillation dominates the flow field in viscoelastic fluids, leading to an early transition into the lock-in condition on a non-dimensional amplitude-frequency plane. 
\section{Conclusions}
In summary, we have numerically investigated the flows of Newtonian and viscoelastic fluids over a circular cylinder oscillating transversely to the streaming fluid at a fixed Reynolds number of $Re = 100$. In particular, the effect of the fluid viscoelasticity on the boundary between the lock-in and no lock-in regions and the underlying vortex structures are analyzed in detail for a range of cylinder oscillation amplitudes and frequencies. Our study has revealed that in the case of a viscoelastic fluid, the lock-in region is found to occur at a low non-dimensional cylinder oscillation frequency at which a no lock-in condition is seen for a Newtonian fluid. Furthermore, a substantial difference in the wake structures is also observed. We report a '2P' (two pairs of vortices in one oscillation cycle) vortex shedding mode for a viscoelastic fluid in the lock-in zone, which does not occur in a Newtonian fluid. This deviation is ascribed to the extra elastic stresses arising from the stretching of the polymer molecules present in viscoelastic fluids. Additionally, the data-driven dynamic mode decomposition (DMD) analysis has been employed to provide a better insight into the competition between the cylinder oscillation frequency and natural vortex shedding frequency in the wake structure. DMD has successfully extracted the coherent flow structures associated with each of these frequencies for both fluids in the lock-in and no lock-in zones. 

\bibliographystyle{jfm}
\bibliography{jfm-instructions}
\end{document}